\documentclass[aps,prl,toolkits,twocolumn]{revtex4}

\usepackage{epsfig,amssymb,amsfonts}

\makeatletter

\renewcommand{\@makefntext}[1]
{\parindent=1em\noindent\hbox to 1.8em{\hss$^{\@thefnmark}$}#1}
\renewcommand{\@footnotemark}{\hbox{\mathsurround=0pt$^{\@thefnmark}$}}

\makeatother

\begin{document}
\title{Comment on ``Parity Doubling and $SU(2)_L \times SU(2)_R$
Restoration in the Hadronic Spectrum"}

\maketitle

In a recent letter Jaffe, Pirjol and Scardicchio (JPS) use
general arguments of chiral symmetry to discuss the possibility
that chiral symmetry might be effectively restored in the hadronic
spectrum \cite{JAFFE} as has been suggested as a possible
explanation of parity multiplets in the hadron spectrum
\cite{G1,CG1,G2,G4,CG2}.  While  the calculations in
ref.~\cite{JAFFE} are correct, the interpretation and discussion
may create the false impression that effective chiral
restoration  (E$\chi$R) in the sense of refs.
\cite{G1,CG1,G2,G4,CG2} is somehow unnatural.  This is
unfortunate; the arguments of JPS are irrelevant to the question
of whether E$\chi$R occurs.

The formal derivation by JPS of the allowable representations is
both correct and quite elegant. However, in terms of the
essential question of whether E$\chi$R can explain parity
doubling, the analysis boils down to the statement that in the
broken phase chiral symmetry {\it alone} does not give rise to
chiral multiplets of degenerate states. This is correct but has
been well known for more than four decades.   Reference
\cite{JAFFE} then notes that some dynamics beyond simple chiral
invariance is required that would suppress those operators that
split ``would be'' chiral partners.  Again this is well known;
indeed it constitutes the underlying idea of refs.
\cite{G1,CG1,G2,G4,CG2}: namely that dynamical effects
responsible for chiral symmetry breaking  become less and less
important in the high-lying hadrons; asymptotically the states
approach a regime where their properties resemble those of an
unbroken phase.  In summary, the JPS analysis recapitulates well
known results without further illuminating the central issue.

Moreover, the presentation of the problem in ref.~\cite{JAFFE}
may create a false impression about the nature of the problem. JPS
emphasize that in the Nambu-Goldstone phase, relations between
masses and couplings characteristic of multiplets in the symmetry
restored phase are possible only if the dynamics suppresses
classes of chirally allowed operators which destroy them.  While
this is correct, it is potentially misleading.  This emphasis
creates the impression that i) if such relations do emerge from
the dynamics they are essentially accidental and {\it
unconnected} with the underlying chiral symmetry and ii) the
emergence of such relations is unnatural since it requires the
fine tuning of many operators. However, both i) and ii) are
fundamentally incorrect.

In point of fact there {\it is} a simple and natural way
connected to chiral symmetry for {\it all} of the operators which
spoil the Wigner-Weyl-like relations to be suppressed
dynamically.   This occurs automatically provided that the states
in question are lying high enough so that they are large
insensitive to the dynamics responsible for spontaneous chiral
symmetry breaking (S$\chi$SB) such as the chiral condensate. As
has been discussed in detail elsewhere there are general
arguments why this may occur naturally high in the spectrum
\cite{CG1,G4}.

The conjecture of  E$\chi$R is precisely  that high lying states
{\it are} largely insensitive to the dynamics of S$\chi$SB,
because they gradually decouple from the quark condensates of the
vacuum, and hence it leads to approximate chiral multiplets. If
this does happen, then numerous unconnected coefficients of the
effective chiral Lagrangian which break chiral symmetry are
suppressed {\it automatically} without additional fine tuning.
Their suppression is profoundly connected to  the underlying
chiral symmetry.

The nonlinear $\sigma$ model upon which JPS base their formalism
obscures this simple fact. In a nonlinear $\sigma$ model
S$\chi$SB is built in from the outset; the underlying dynamics is
not explicit and these effects are encoded in the coefficients of
various tree-level operators.  If one starts from such a model
and then imposes relations characteristic of the unbroken phase
for some class of states, it appears that one is demanding that
the coefficients multiplying many chirally allowed independent
operators are simultaneously suppressed.  This may seem to demand
a very unlikely conspiracy among the coefficients.  However,
these coefficients are encoding the effects of the dynamics of
S$\chi$SB.  If the state in question is  insensitive to the
chiral symmetry breaking condensates,  all the coefficients are
automatically suppressed.

Reference \cite{JAFFE} is also misleading in another way.  The
concluding sentence concerns explanations for parity multiplets
``other than restoration of $SU(2)_L \times SU(2)_R$ which as we
have shown cannot occur.''  The reader may take this to mean that
the JPS argument has somehow ruled E$\chi$R. This is emphatically
not the case.

{Thomas D.~Cohen${}^1$   and Leonid Ya. Glozman${}^2$} \\ \\
 {${}^1$Department of Physics, University of
Maryland, College Park, MD 20742, USA}\\{${}^2$Institute for
Physics, University of Graz, Universit\"atsplatz 5, A-8010 Graz,
Austria}

\end{document}